# Monolayer graphene panorama, Majorana modes and Longitudinal conductivity

PARTHA GOSWAMI

*Deshbandhu College, University of Delhi, Kalkaji, New Delhi-110019,India*

E-mail: physicsgoswami@gmail.com;Tel:0091-129-243-9099.

**ABSTRACT** We take a wide-angle view of the problem of monolayer graphene where the valley-mixing and the spin-degeneracy lifting are assumed to be possible by wedging in the requisite ingredients, viz. the atomically sharp scatterers and the strong Rashba coupling dominating over the intrinsic spin-orbit coupling. This leads to eight Majorana-like modes (quasi-particles which are self-conjugate) close to the experimentally inaccessible Dirac points. Using Kubo formula we also show that the semi-classical diffusive (longitudinal) conductivity is nearly (2.018 $e^2/h$) at room temperature for the disordered system. Though this is an overestimation, we have been, never-the-less, able to qualitatively capture the fact that the room temperature conductivity of graphene is finite and the contribution to the conductivity arises from the momentum very close to the Dirac points.

**1. Introduction** Ever since Fu and Kane have predicted [1] a one-dimensional mode of Majorana fermions (half-integer-spin (relativistic) particles which are their own anti-particles) at the interface between a conventional super-conductor and a superconducting topological insulator(TI) surface state, there has been persistent effort[2,3,4] to obtain signature of this elusive mode within such systems. The Netherland group of Mourik et al.[5] has been crowned with success recently in this endeavor. They have reported the realization of proximity-induced topological super-conductivity and the formation of Majorana bound states in the Indium antimonide (InSb) quantum wires. The remarkable discovery of the Netherland group [5] also agrees with the more recent theoretical works [6,7]of Sato et al. and Sau et. al. where the latter, for example, have demonstrated that a topological superconducting phase could be realized using a semi-conductor quantum well coupled to an s-wave superconductor and a ferromagnetic insulator. To understand how Majorana resonance comes about in a quantum wire system we note that, in general, inside a crystal the conducting electrons have their counterpart as mobile "holes" which are formed when an electron moves out of a stable site in the crystal lattice. In order that electron and hole preserve their individual identities in a electron-hole mixture (usually a free electron annihilates a hole and both disappear), one combines a superconductor with a topological insulator (the latter conducts electricity only on its surface). The two materials in conjunction create a pattern of electric fields at their boundary which can stop electrons-hole annihilation potentially allowing Majorana fermions to form.

The InSb nanowire, used by Mourik et al.[5] in their experiment, bridged the gap between a superconducting electrode made of niobium titanium nitride (NbTiN) and a normal electrode made of gold. The device is cooled to temperature T~ 50 mK and a magnetic field is applied along the direction of the nanowire. The current flowing through the nanowire was measured as a function of voltage. At zero applied magnetic field, two small peaks in the conductance were observed on either side of zero applied voltage. When the applied magnetic field was increased, these peaks remained in the same position. This also occurred when an electric field was applied to the nanowire. This absence of response by the peaks to magnetic and electric fields was explained by the presence of pairs of Majorana fermions (MFs) at one end of the nanowire. A normal metal (gold) contact as a tunnel probe was used in order to exclude super-current as an explanation for the zero bias peak. The MFs could be important for the storage and transmission quantum information because these species, unlike the usual"Dirac" fermions, obey "non-Abelian statistics" and therefore should be resistant to environmental noise. In other words, a qubit encoded in the Majorana pair is expected to have an unusually long coherence time. The team of Mourik et al.[5], however, are yet to establish the non-Abelian nature of the resonance modes they discovered. In the backdrop of the excitement generated due to this finding, it is useful to review how such modes could be constructed theoretically in a host Dirac system, viz. graphene, in the 1+2 space-time dimensions. It may be noted that the Majorana modes were originally predicted by E. Majorana[8] nearly seventy years ago in 1+3 space-time dimensions.

In this communication we consider the kinetic term in the single particle Hamiltonian ($H_0$) in real space of the Dirac systems, which may be represented in a compact form by $H_0 = -i\hbar v_F (\tau_3 \otimes \sigma_1 \partial_x + \tau_3 \otimes \sigma_2 \partial_y)$, where $\tau_{0,1,2,3}$ ($\tau_0 = I_{2X2}$(identity matrix), $\tau_1 = \tau_x$, $\tau_2 = \tau_y$, $\tau_3 = \tau_z$) and $\sigma_{0,1,2,3}$ ($\sigma_0 = I_{2X2}$, $\sigma_1 = \sigma_x$, $\sigma_2 = \sigma_y$, $\sigma_3 = \sigma_z$) are two independent sets of 2 × 2 Pauli matrices. In graphene, the Pauli matrices $\tau_{0,1,2,3}$ correspond to the **K** and **K′** valley (iso-spin) index whereas the Pauli matrices $\sigma_{0,1,2,3}$ correspond to the A and B sub-lattice (pseudo-spin) index. The Dirac equation with this Hamiltonian is given by $H_0 \Psi = \varepsilon \Psi$ where

$$\Psi = \begin{pmatrix} \psi^B_K \\ \psi^A_K \\ \psi^B_{K'} \\ \psi^A_{K'} \end{pmatrix}. \qquad (1)$$

is a four-component spinor. An additional mass($m(\mathbf{r})$)term in the Dirac equation in graphene can be viewed as a bosonic field (an order parameter) generated due to the spontaneous breaking of a symmetry, such as the chiral symmetry. This may be dubbed as the Higgs mechanism in the Dirac systems in 1+2 -space-time dimensions. There could be topological defects (TD) in the order parameter as well, such as vortices [9] in a type–II superconductor. In graphene, the mass order parameter could be induced, for example, by placing the system on a certain substrate where there is a difference in the potential [10], seen by the two atoms in the unit cell of graphene, which creates a charge-density wave (CDW) gap with broken chiral symmetry. The single-particle excitation spectrum, however, is particle-hole symmetric and, therefore, "hiding" the charge difference one may construct the Majorana modes out of electron and hole excitations provided one has access to a situation characterized by broken iso-spin symmetry and spin non-degeneracy[11]. We find them as the necessary conditions required for the construction of Majorana modes in graphene. The iso-spin symmetry breaking is possible if the CDW gap generating potential corresponds to atomically sharp scatterers. To explain, we quote here that Suzuura et al. [12,13,14] in a different context have suggested several years ago that, when the inter-valley scattering rate is higher than the de-coherence rate, the inter-valley particle–particle correlation function(PPCF) is enhanced leading to a conventional weak localization(WL). These authors have reported that, even in the absence of spin-orbit coupling, from the possible weak anti-localization(WAL–positive magneto-resistance beyond a critical magnetic field $B_i$) a WL(negative magneto-resistance for all possible magnetic field strength) may be obtained by a strong inter-valley scattering from the atomically sharp scatterers, while the crossovers from the latter to the former are obtained by reducing the disorder strength down to the ballistic limit [14]. In addition, it was shown that the trigonal warping inclusion in the monolayer graphene Hamiltonian [14,15] suppresses the intra-valley PPCF and, therefore, WAL as well in the case when electrons do not change their valley state; the inter-valley PPCF is not affected by trigonal warping in the case of weak inter-valley scattering due to the time-reversal invariance of the system. In view of these published results we visualize (do not visualize) a major role of the inter-valley scattering(the trigonal warping) in the search Majoranas in graphene. The preliminary elucidation of the requirements for showing Majoranas in graphene is the primary goal of this communication. For example, it is well-known that the traversal of a closed contour in momentum space, corresponding to the rotation of electronic momentum δ**k** by angle θ = 2π around the Dirac point, the chiral wave function in mono-layer graphene adiabatically evolves to undergo a phase change of π known as Berry's phase (arising from the rotation of the pseudo-spin degree of freedom). In the construction of the Majorana modes out of electron and hole excitations, we find that one requires θ = 4π, 8π,….. to hide this phase change. Despite all these, we note that there is no way to trap these modes at this moment for the demonstration purpose in our scheme. We add that there are two in-equivalent representations for real Dirac γ-matrices (see section 2), which obey the anti-commutation rule of Clifford algebra [16], corresponding to the valleys **K** and **K** ′. As has been paraphrased in ref.[17], the direct sum of these irreducible representations corresponds to a tensor product space (shown above). The latter does not disfavor the valley-mixing in odd space-time dimensions. The issue is not pertinent for the Majorana nanowire of Mourik et al. [5] as TI has a single Dirac cone. It is therefore quite likely that their zero-bias conductance peaks have have their origin in the weak anti-localization in the Majorana nanowire [18].

Graphene has weak intrinsic spin-orbit interaction(15- 30 meV) as the carbon nuclei is light and weak hyperfine coupling as carbon materials consist predominantly of the nuclear spin free $^{12}$C isotope. This makes it potentially a good spin conductor with long spin coherence times [19,20]. In section 3 we shall discuss in brief how the strong Rashba spin-orbit (tunable by gate voltage) leads to the spin-degeneracy lifting which is shown to be yet another requirement for the existence of Majoranas in graphene.

Alternative to the complex scenario portrayed in the paragraph before the last one is that the substrate could be a superconductor leading to a particle-hole symmetric excitation spectrum. Moving on with this choice, we recall that the states in a Dirac system at different energy (that is states from the valence band and from the conduction band), in general, arise from the different valleys **K** and **K**′. Once again, one is interested here in zero-energy (ε = 0) mid-gap, real solutions (Majorana-like quasi-particles) which should be localized/ quasi-localized in space. We note that in order to localize such states one needs to have TDs, such as quantum vortices [9], which can trap the "so-called" mid-gap zero modes(the topological protection of these modes is guaranteed by the Atiyah-Singer index theorem[21,22]). To pave the ground to include the superconducting order parameters (together with TDs as vortices) in the Dirac equation, one must introduce one more grading relating to particle-hole (represented by the Pauli matrices $\mu_{0,1,2,3}$($\mu_0 = I_{2X2}$, $\mu_1 = \mu_x$, $\mu_2 = \mu_y$, $\mu_3 = \mu_z$)) in (1). Thus, including a vector potential **A**= ($A_x$, $A_y$ )( equivalently, **A**= −**e**$_\theta$A(r), where **e**$_\theta$ =(−sin(θ),cos(θ)), in the plane polar coordinate system), one may now write the full Hamiltonian in compact form as H = [$v_F$ ($\tau_3 \otimes \mu_0 \otimes \sigma_1$ Л$_x$ + $\tau_3 \otimes \mu_3 \otimes \sigma_2$ Л$_y$) + $m(\mathbf{r}) \otimes \mu_3 \otimes \sigma_0$] where the operators Л$_x$≡ − $i$ ℏ $\partial_x$ −q $A_x$ , Л$_y$ ≡ − $i$ ℏ $\partial_y$ −q $A_y$, and the matrix

$$m(\mathbf{r}) = \begin{bmatrix} 0 & \Delta(r) \\ \Delta^*(r) & 0 \end{bmatrix}. \quad (2)$$

To introduce real spin, of course, yet another grading (represented by the Pauli matrices $s_{0,1,2,3}$) must be inserted which we shall discuss in section4. The Hamiltonian H is the low-energy Ghaemi-Wilczek [23] version of the Dirac Hamiltonian with the usual Peierls substitution including the superconducting order parameter Δ(r). The pair potential corresponds to opposite 2-momentum spin-singlet states requiring inter-valley mixing [23].

With this type of pairing, the authors[23] could show the existence of quasi-localized near zero modes in graphene.

The paper is organized as follows: In section 2 we identify real (imaginary) Dirac matrices for Dirac( Majorana) fermions so that the longitudinal ($\sigma_{xx}$) conductivity could be calculated using them in the Kubo formula[24] as velocity operators. In section 3 we obtain Majorana operators in terms of Dirac creation and annihilation operators of second quantization. We calculate $\sigma_{xx}$ ( as well as the Hall conductivity) in section 4. The paper ends with some concluding remarks in section 5.

**2. Dirac matrices as velocity operators** In this section our aim is to identify real (imaginary) Dirac matrices for Dirac( Majorana) fermions so that they could be used as the velocity operators in the calculation of longitudinal conductivity in the section 4. We also find here that the signature of the Lorentz metric for Majoranas is different from that of Dirac fermions; the two metrics could be degenerate in some hyper-space. To carry out these tasks, we consider the low-energy Dirac Hamiltonian $H_0$ appearing above Eq.(1). It may be written as $H_0 = \hbar v_F (\xi \delta k_x \sigma_x + \delta k_y \sigma_y) = v_F (\xi p_x \sigma_x + p_y \sigma_y)$ where the valley index $\xi = \pm$ denotes the two Dirac points **K** and **K**′, respectively. Introducing the notation $\sigma^1 \equiv \sigma_x$, $\sigma^2 \equiv \sigma_y$, $p^1 \equiv p_x$, $x^1 \equiv x$, and $p^0 \equiv i p_y$, we may write $H_0 = v_F (\xi \sigma^1 p^1 + \sigma^2 p^0/i)$. Furthermore, we identify $p^0 = -i\hbar(\partial/\partial x^0) = -i\hbar \partial_0$ (which implies that $x^0 = -i\, y$) where $\partial_0 \equiv (\partial/\partial x^0)$. We also introduce the counterpart, of the Dirac matrices in 1+3 space-time dimensions, in 1+2 space-time dimensions as $\gamma^0 \equiv -i\sigma^2$, $\gamma^1 \equiv \xi \sigma^1$, and $\gamma^2 \equiv \sigma^3$. This allows us to write $H_0 = v_F (\gamma^1 p^1 + \gamma^0 p^0) = -i\hbar v_F (\gamma^1 \partial_1 + \gamma^0 \partial_0)$. Since the energy eigenvalues of the Hamiltonian are $E = \pm v_F |\mathbf{p}|$, i.e. $p_x^2 + p_y^2 = (E/v_F)^2$, it is easy to see that $p^2 = \pm (E/v_F)$ yielding the operator $p^2 = \pm(i\hbar/v_F)(\partial/\partial t) = \pm (i\hbar) \partial_2$ with $\partial_2 \equiv (1/v_F)(\partial/\partial t)$ which gives us $x^2 = \mp i\, v_F\, t$. In view of the time-dependent Dirac equation $i\hbar v_F \gamma^2 \partial_2 \psi = H_0 \psi$, where $\psi$ is a two-component spinor, the Dirac equation in (1+2) dimensions for mass-less real fermions in covariant form may be written as $i(\gamma^\mu \partial_\mu)\psi = 0$. We find that $\gamma^0 \gamma^0 = - I_{2X2}$ (identity matrix) and $\gamma^1 \gamma^1 = \gamma^2 \gamma^2 = I_{2X2}$ and all three matrices $\gamma^0$, $\gamma^1$, and $\gamma^2$ anti-commute in pairs. Thus, they comply with the usual rules of Clifford algebra[17] in Minkowski space. In this three-component description, since the real matrices $\{\gamma^0, \gamma^1, \gamma^2\}$ render the Hamiltonian real, the full, space-time dependent field $\psi$ is complex. In the absence of a mass term, the kinetic term $\int dt dx dy\, \bar\psi(i\gamma^\mu \partial_\mu \psi)$ in the 1+2 dimensional fermion action remains invariant if one implements a parity transformation by making the replacement $\psi(x, y, t) \rightarrow \gamma^1 \psi(-x, y, t)$ and change the integration variables accordingly; a hypothetical mass term of the form $(m \int dt dx dy\, \psi^\dagger \gamma^0 \psi)$ in the fermion action changes sign under parity. We define the Dirac adjoint for any bi-spinor matrix M or spinor as follows: $\bar M \equiv \gamma^0 M^\dagger \gamma^0$, $\bar\psi \equiv \psi^\dagger \gamma^0$, $\bar\gamma^\alpha = \gamma^\alpha$. Note that $\bar\psi$ is a row vector, while $\psi$ is a column vector. Since $\bar\gamma^\alpha = \gamma^\alpha$, we may say that the gamma matrices are Dirac Hermitian. The anti-commutation and Hermiticity rules are the only constraints on the gamma matrices, so any set of matrices satisfying these constraints is a valid representation. The mass(m) term that could be added to open a spectral gap are proportional to either $\gamma^2 \gamma^2$ or $\gamma^2$. The former is even under parity, but odd under time reversal (which interchanges **K** and **K**′(see section 3)). The latter is odd under parity (which interchanges the A and B sub-lattices). With the former (latter) one may write the covariant equation as $(i\gamma^\mu \partial_\mu - m\, \gamma^2 \gamma^2)\psi = 0$ $((i\gamma^\mu \partial_\mu - m\, \gamma^2)\psi = 0)$. The field $\psi$ here corresponds to a massive Dirac fermion (or a charge-non-self-conjugate fermion). The complex conjugate of $\psi$, denoted by $\psi^*$, respectively, satisfies the equations $(i\gamma^\mu \partial_\mu - m\, \gamma^0 \gamma^0)\psi^* = 0$ and $(i\gamma^\mu \partial_\mu + m\, \gamma^2)\psi^* = 0$.

We notice from above that the square of the "infinitesimal physical distance" $ds^2 = (dx^0)^2 + (dx^1)^2 + (dx^2)^2 = (dx)^2 - (dy)^2 - (v_F \, dt)^2$. This yields the space-time modeled by a pseudo-Riemannian manifold and the relevant Lorentz metric as

$$\eta_{\alpha\beta} = \begin{pmatrix} 1 & 0 & 0 \\ 0 & -1 & 0 \\ 0 & 0 & -1 \end{pmatrix}. \tag{3}$$

The indices α,β run over 0,1,2 with $x^2$ as the time coordinate and ($x^0$, $x^1$) as the space coordinates. The signature of a metric refers to how many time-like and space-like characters are in the space-time; the metric is positive definite on the space-like subspace, and negative definite on the time-like subspace. We find that it is positive definite in the *x* direction, and negative definite in the *y* direction and in the time direction. We also find that in 3 space-time dimensions there exist two in-equivalent representations for real γ-matrices ( which is true for any odd number of space-time dimensions [17]):$\gamma^0 = -i\sigma^2$, $\gamma^1 = \xi\,\sigma^1$, $\gamma^2 = \sigma^3$ (where $\xi = \pm 1$). One may use the first of these representations for the expansion around the Dirac point **K** and the second one for the point **K′**. The direct sum [17] of these irreducible representations corresponds to the tensor product space discussed in section 1. We may now consider the Ehrenfest theorem also valid in relativistic quantum mechanics. According to the theorem,$(d/dt)\langle \boldsymbol{O} \rangle = \langle d\boldsymbol{O}/dt \rangle = (1/i\hbar)\langle [\boldsymbol{O}, H_{Dirac}] \rangle$, where $H_{Dirac} = (i\gamma^\mu \partial_\mu - m\,\gamma^2)\,\psi = 0$ and the conical brackets refer to the average calculated with the eigenstates of this Hamiltonian. The average of the velocity operator $(d/dt)\langle \boldsymbol{r} \rangle = \langle d\boldsymbol{r}/dt \rangle = (1/i\hbar)\langle [\boldsymbol{r}, H_{Dirac}] \rangle$ where *r* is the position operator in coordinate representation. Upon comparing graphene with 1+3 QED system, it is easy to see that the velocity operator $\frac{dr}{dt}$, in coordinate/momentum representation, for the former is $v_F$ ($\gamma^1 = \xi\,\sigma^1$, $\gamma^0 = -i\sigma^2$). This one-particle operator definition does not mix the solutions arising out of the valleys **K** and **K′**.

We now consider a scenario where the field ψ is real and corresponds to a charge-self-conjugate Dirac fermion (or Majorana fermion) satisfying a Dirac equation involving imaginary γ-matrices. Accordingly, one chooses a different representation for such matrices, viz. $\gamma^0 = \sigma^2$, $\gamma^1 = i\sigma^3$, $\gamma^2 = i\xi\sigma^1$ where all the matrices are imaginary. We may note that, with this representation, $\gamma^0\gamma^0 = I_{2X2}$(identity matrix), $\gamma^1\gamma^1 = = \gamma^2\gamma^2 = -I_{2X2}$(identity matrix) and all three matrices $\gamma^0$, $\gamma^1$, and $\gamma^2$ anti-commuting in pairs. Thus, they comply with the usual rules of Clifford algebra. Introducing the notation $\sigma^1 \equiv \sigma_x$, $\sigma^2 \equiv \sigma_y$, $ip^2 \equiv p_x$ and $p^2 = -i\hbar\,\partial_2$ where $\partial_2 \equiv (\partial/\partial x^2)$ (which implies that $x^2 = i$ x), and $p^0 \equiv p_y = (-i\hbar)(\partial/\partial y) = (-i\hbar)\,\partial_0$ (where $\partial_0 \equiv (\partial/\partial x^0)$ which implies that $x^0 = y$) we may write the Dirac Hamiltonian as $H_D = -i\,\hbar v_F(\gamma^2\partial_2 + \gamma^0\partial_0)$. Since the energy eigenvalues of the Hamiltonian are $E = \pm v_F\,|\boldsymbol{p}|$, i.e. $p_x^2 + p_y^2 = (E/v_F)^2$, we identify that $p^1 = \pm(E/v_F)$ yielding the operator $p^1 = \pm(i\hbar/v_F)(\partial/\partial t) = \pm i\hbar\,\partial_1$ (with $\partial_1 \equiv (1/v_F)(\partial/\partial t)$). This implies that $x^1 = \mp i\,v_F\,t$. Therefore, the square of the "infinitesimal physical distance" $ds^2 = (dx^0)^2 + (dx^1)^2 + (dx^2)^2 = -(dx)^2 + (dy)^2 - (v_F\,dt)^2$ which yields the relevant Lorentz metric as

$$\eta_{\alpha\beta} = \begin{pmatrix} -1 & 0 & 0 \\ 0 & 1 & 0 \\ 0 & 0 & -1 \end{pmatrix}. \tag{4}$$

The indices α,β run over 0,1,2 with $x^1 = \mp i\, v_F t$ as the time coordinate and $(x^0, x^2)$ as the space coordinates. We find that the signature of the Lorentz metric here (positive definite in the *y* direction, and negative definite in the *x* direction and in the time direction) is different compared to the Dirac fermion case. Thus, the two metrics are degenerate on some hyper-surfaces. In view of the time-dependent Dirac equation $i\hbar(\partial/\partial t)\,\psi = H_D\,\psi$, where ψ is a two-component spinor, and $\partial_1 \equiv (1/v_F)(\partial/\partial t)$, the Dirac equation in (1+2) dimensions for a massless fermion in covariant form may be written as $i\gamma^1 \partial_1 \psi = -i(\gamma^0 \partial_0 + \gamma^2 \partial_2)\psi$, or $i(\gamma^\mu \partial_\mu)\psi = 0$. In the presence of a mass term proportional to $I_{2X2}$, this equation will appear as $(i\gamma^\mu \partial_\mu + m\,\gamma^1\gamma^1)\psi = 0$ whereas, for the mass term proportional to $i\gamma^1$, we have $(i\gamma^\mu \partial_\mu + i\,m\,\gamma^1)\psi = 0$. Since $\gamma^\mu$ are imaginary matrices, it is easy to see that the complex conjugate of the space-time dependent fields ψ in these equations satisfy the same equations. Thus, the field ψ here is real and it corresponds to a massive Dirac fermion which is charge-self-conjugate and known as Majorana fermions. We once again find that in 1+2 space-time dimensions there exist two in-equivalent representations for complex γ-matrices: $\gamma^0 = \sigma^2$, $\gamma^1 = i\sigma^3$, $\gamma^2 = i\xi\sigma^1$ (where ξ = ±1). We shall use the first of these representations for the expansion around the Dirac point **K** and the second one for the point **K′**. It is now easy to see that the velocity operator for Majoranas is $v_F(\gamma^2 = i\xi\sigma^1, \gamma^0 = \sigma^2)$.

**3. Majorana operators in terms of Dirac creation and annihilation operators of second quantization** The discussion in the previous section clearly indicates that the Majoranas and the Dirac fermions in a 1+2 space-time system are intimately connected. The main issue from a theoretical perspective is to find the condition(s) under which the former could be realized in such systems. To look for the condition(s) in graphene, we shall express mathematically the Dirac and Majorana fermions in terms of the creation and annihilation operators of second quantization in this section. Our expressions will encode the electron's and hole's characteristic fermion statistics, the particle–antiparticle correspondence, and the unusual anti-commutation relations obeyed by the Majorana operators in a transparent manner. Before taking up this task we explain in brief what type of excitations we are looking for. The 'particle states' are associated with creation operators $d_j^\dagger$, and 'antiparticle (hole) states' with their conjugate operators, $d_j$. This means $d_j^\dagger$ can create a particle, or destroy a hole in a state j, whereas $d_j$ can create a hole, or destroy a particle, in a state j. For two distinct (orthogonal) states j and k, the anti-symmetry of Fermi-Dirac statistics implies that $\{d_j^\dagger, d_k^\dagger\} = \{d_j, d_k^\dagger\} = \{d_j, d_k\} = 0$. The completeness relation, on the other hand, implies that $\{d_j, d_j^\dagger\} = 1$. We notice that the particle–hole interchange (charge conjugation) is implementable by $d_j \leftrightarrow d_j^\dagger$. Thus, for a Dirac fermion the operators $d_j^\dagger$ and $d_j$ are distinct, while for a Majorana fermion they are identical. One example, which complies apparently with the implementation, is an exciton (bound states of electron and hole). In the language of second quantization, this will correspond to $\hat{A}_{exciton} \equiv (d_j d_k^\dagger + d_j^\dagger d_k)$. Obviously enough, under charge conjugation, the exciton 'creation' operator $\hat{A}_{exciton}$ goes over to itself, and the concomitant excitations are their own antiparticles. But excitons are always bosons, with integer spin, and thus could not be Majoranas. Our candidate, therefore, is an operator $\hat{C}_j = d_j^\dagger + d_j$ which corresponds to a particle and hole mixture creation in a state j. It will hide the 'charge' completely without tinkering with the spin.

We now turn our attention to a monolayer graphene Hamiltonian (1). This will keep our description at the easily comprehensible level. In the second quantized language, introducing the real spin σ, we may write

$$H_0 = \sum_{\delta \mathbf{k}} \xi \hbar v_F |\delta \mathbf{k}| \; (a^\dagger_{\delta \mathbf{k},\sigma} \; b^\dagger_{\delta \mathbf{k},\sigma}) \begin{bmatrix} 0 & \exp(-i\xi\theta_k) \\ \exp(i\xi\theta_k) & 0 \end{bmatrix} \begin{pmatrix} a_{\delta \mathbf{k},\sigma} \\ b_{\delta \mathbf{k},\sigma} \end{pmatrix} \quad (5)$$

where $\cos(\theta_k) = \delta k_x / |\delta \mathbf{k}|$, $\sin(\theta_k) = \delta k_y / |\delta \mathbf{k}|$, and $\theta_k = \arctan(\delta k_y / \delta k_x)$. The operators $a^\dagger_{\delta k,\sigma}$ and $b^\dagger_{\delta k,\sigma}$ respectively, correspond to the fermion creation operators for A and B sub-attices in the monolayer. The matrices $H_{0,\mathbf{K}}$ and $H_{0,\mathbf{K}'}$ for $\mathbf{K}$ and $\mathbf{K}'$, respectively, have two eigenvalues $\pm \hbar v_F |\delta \mathbf{k}|$. One may write the normalized eigenfunctions, in momentum space, for the momentum around $\mathbf{K}$ and $\mathbf{K}'$ in the compact form as

$$\psi_{\pm, \mathbf{K}}(\delta \mathbf{k}) = (1/\sqrt{2}) \begin{bmatrix} \exp(-i\theta_k/2) \\ \pm \exp(i\theta_k/2) \end{bmatrix}, \; \psi_{\pm, \mathbf{K}'}(\delta \mathbf{k}) = (1/\sqrt{2}) \begin{bmatrix} \exp(i\theta_k/2) \\ \mp \exp(-i\theta_k/2) \end{bmatrix}, \quad (6)$$

We have $+\hbar v_F |\delta \mathbf{k}| \rightarrow \psi_{+, \mathbf{K}(\mathbf{K}')}(\delta \mathbf{k})$ (conduction band(electron state)) and $-\hbar v_F |\delta \mathbf{k}| \rightarrow \psi_{-, \mathbf{K}(\mathbf{K}')}(\delta \mathbf{k})$ (valence band(hole state)). One notices that $\psi_{+,\mathbf{K}}$ is complex conjugate of $\psi_{-,\mathbf{K}'}$ and $\psi_{-,\mathbf{K}}$ is complex conjugate of $\psi_{+,\mathbf{K}'}$. We make the identification that $\psi_{c, \mathbf{K}(\mathbf{K}')}(\delta \mathbf{k}) = \psi_{+, \mathbf{K}(\mathbf{K}')}(\delta \mathbf{k})$ and $\psi_{v, \mathbf{K}(\mathbf{K}')}(\delta \mathbf{k}) = \psi_{-, \mathbf{K}(\mathbf{K}')}(\delta \mathbf{k})$ where the subscript c(v) refers to the conduction(valence) band. We find that $\psi_{c(v), \mathbf{K}}(\delta \mathbf{k}) = (1/\sqrt{2})(|1\rangle_\mathbf{K} \pm |0\rangle_\mathbf{K})$, and $\psi_{c(v), \mathbf{K}'}(\delta \mathbf{k}) = (1/\sqrt{2})(|1\rangle_{\mathbf{K}'} \mp |0\rangle_{\mathbf{K}'})$, where the upper(lower) sign corresponds to the subscript c(v). Here $|1\rangle_{\mathbf{K}(\mathbf{K}')} = \exp(\mp i\theta_k/2) |\uparrow\rangle$ and $|0\rangle_{\mathbf{K}(\mathbf{K}')} = \exp(\pm i\theta_k/2) |\downarrow\rangle$; in the states $|1\rangle_{\mathbf{K}(\mathbf{K}')}$ and $|0\rangle_{\mathbf{K}(\mathbf{K}')}$ the upper(lower) sign corresponds to the subscript $\mathbf{K}(\mathbf{K}')$. Here $|\uparrow\rangle = \begin{bmatrix} 1 \\ 0 \end{bmatrix}$ and $|\downarrow\rangle = \begin{bmatrix} 0 \\ 1 \end{bmatrix}$, respectively, are the 'so-called' up and down states arising out of the choice of the sub-lattice basis (A, B). We, therefore, notice that the existence of two independent sub-lattices A and B(corresponding to the two atoms per unit cell) leads to the existence of novelty in graphene dynamics where the two linear branches of graphene energy dispersion (intersecting at Dirac points) become independent of each other, indicating the existence of a pseudo-spin quantum number analogous to electron spin (but completely independent of real spin). In other words, the existence of the pseudo-spin quantum number is a natural byproduct of the basic lattice structure of graphene comprising two independent sub-lattices. The eigenstates above in the vicinity of the $\mathbf{K}$ and $\mathbf{K}'$ points necessitate the introduction of the notion of iso-spin, once again reminiscent of the states of the spin-1/2 operator.

To clarify the notion of the iso-spin, we note that $\psi_{\pm, \mathbf{K}}(\delta \mathbf{k})$ and $\psi_{\pm, \mathbf{K}'}(\delta \mathbf{k})$ are linked by a symmetry property provided we establish the correspondence between the states around the valleys $\mathbf{K}$ and $\mathbf{K}'$ with real single spin-1/2 operator. For this we draw the analogy with a single spin-1/2 operator $\mathbf{S}$ represented in terms of Pauli matrices $\sigma^i$: $S^i = (1/2)\hbar \sigma^i$. The eigenvalues of $\sigma^i$ are $\pm 1$ and the corresponding eigenstates of $\sigma^z$, say, are $|\uparrow\rangle = \begin{bmatrix} 1 \\ 0 \end{bmatrix}$, and $|\downarrow\rangle = \begin{bmatrix} 0 \\ 1 \end{bmatrix}$. The operator $S^i$ in the second quantized language can be written as $\mathbf{S}^i = \sum_{\mu,\mu'} d^\dagger_\mu S^i_{\mu,\mu'} d_{\mu'}$ where $d^\dagger_\mu$ creates a particle in the state $|\mu\rangle$. This immediately gives $S^x = (1/2)(d^\dagger_\uparrow d_\downarrow + d^\dagger_\downarrow d_\uparrow)$, $S^y = (1/2i)(d^\dagger_\uparrow d_\downarrow - d^\dagger_\downarrow d_\uparrow)$, and $S^z = (1/2)(d^\dagger_\uparrow d_\uparrow - d^\dagger_\downarrow d_\downarrow)$. The spin-reversal operators are $S^+ = d^\dagger_\uparrow d_\downarrow$ and $S^- = d^\dagger_\downarrow d_\uparrow$. The anti-unitary time reversal operator for real spins is defined as $\hat{A} = \Theta$ κ where $\Theta = \exp(i\pi S^y/\hbar)$, and κ is the complex conjugation operator. The operator Θ(an orthogonal matrix) is given by

$$\begin{bmatrix} 0 & 1 \\ -1 & 0 \end{bmatrix}.$$

One may then write $\hat{A}\,\psi_\uparrow = \psi^*_\downarrow$ and $\hat{A}\,\psi_\downarrow = -\psi^*_\uparrow$. Having accomplished this exercise, we notice that analogously $\hat{A}\,\psi_{+,K}(\delta\mathbf{k}) = \psi_{-,K}(\delta\mathbf{k}) = \psi^*_{+,K'}$ and $\hat{A}\,\psi_{-,K}(\delta\mathbf{k}) = -\psi_{+,K}(\delta\mathbf{k}) = -\psi^*_{-,K'}$. In fact, we also notice from above that $\hat{A}\,\Theta^{-1} = I$, $\hat{A}\,H_K = -H_K\Theta$ and $\hat{A}\,H_{K'} = -H_{K'}\Theta$ which yield $\Theta\kappa H_K\Theta^{-1} = -H_K = H^*_{K'}$, or, $\Theta H^*_K\Theta^{-1} = H^*_{K'}$. The difference, however, is that whereas $\psi_\uparrow$ and $\psi_\downarrow$ are spinors $\begin{bmatrix} 1 \\ 0 \end{bmatrix}$ and $\begin{bmatrix} 0 \\ 1 \end{bmatrix}$ with real elements, $\psi_{+,K}(\delta\mathbf{k})$, $\psi_{+,K'}$, etc. are Dirac spinors with complex elements. None-the-less, upon assuming that the valley states somehow correspond to real spins we find that states around **K** and **K'** are linked by a symmetry akin to the time-reversal symmetry of real spins. In graphene, electronic density is usually shared between A and B sub-lattices, so that an iso-spin indexed wave function is a linear combination of 'up' and 'down' as shown above. We see that not only do the electrons possess the iso-spin degree of freedom, but they are chiral, meaning the orientation of the pseudo-spin **σ** is related to the direction of the electronic momentum **p**. We introduce the chirality (helicity) for the system to characterize the eigenfunctions through the projection of the momentum operator along the direction of the operator $\boldsymbol{\sigma} = (\sigma_x, \sigma_y)$ (or $\boldsymbol{\sigma}^* = (-\sigma_x, \sigma_y)$). The chirality operator is defined as $\hat{C} = (1/2)\,\boldsymbol{\sigma}\cdot\dfrac{\delta\mathbf{k}}{|\delta\mathbf{k}|}$ for momentum around **K** and $\hat{C}^* = (1/2)\,\boldsymbol{\sigma}^*\cdot\dfrac{\delta\mathbf{k}}{|\delta\mathbf{k}|}$ for momentum around **K'**. The chirality operators $\hat{C}$ and $\hat{C}^*$, respectively, commutes with $H_{0,K}$ and $H_{0,K'}$. We, thus, find that the chirality corresponds to a good quantum number around **K** and **K'**.

The stage is now set to introduce the Majorana-like operators in the second quantization language for graphene. In a bid to achieve this, we first recall that the operators $a^\dagger_{\delta\mathbf{k},\sigma}$ and $b^\dagger_{\delta\mathbf{k},\sigma}$ with momentum $\delta\mathbf{k}$ and spin σ, respectively, have been used for the fermion creation operators for A and B sub-lattices. Suppose the creation operators for c(v) be denoted by $d^\dagger_{c,\delta\mathbf{k},\sigma}$ ($d^\dagger_{v,\delta\mathbf{k},\sigma}$). In view of above, around **K** we may define

$$d^\dagger_{c(v),\,\delta\mathbf{k},\sigma}(\mathbf{K}) = (1/\sqrt{2})a^\dagger_{\delta\mathbf{k},\sigma}(\mathbf{K})\exp(-i\theta_k/2) \pm (1/\sqrt{2})b^\dagger_{\delta\mathbf{k},\sigma}(\mathbf{K})\exp(+i\theta_k/2). \quad (7)$$

This leads to $a^\dagger_{\delta\mathbf{k},\sigma}(\mathbf{K}) = (1/\sqrt{2})\exp(i\theta_k/2)(d^\dagger_{c,\delta\mathbf{k},\sigma}(\mathbf{K}) + d^\dagger_{v,\delta\mathbf{k},\sigma}(\mathbf{K}))$, and $b^\dagger_{\delta\mathbf{k},\sigma}(\mathbf{K}) = (1/\sqrt{2})\exp(-i\theta_k/2)(d^\dagger_{c,\delta\mathbf{k},\sigma}(\mathbf{K}) - d^\dagger_{v,\delta\mathbf{k},\sigma}(\mathbf{K}))$. Similarly, around **K'**, we may define

$$d^\dagger_{c(v),\,\delta\mathbf{k},\sigma}(\mathbf{K'}) = (i/\sqrt{2})a^\dagger_{\delta\mathbf{k},\sigma}(\mathbf{K'})\exp(i\theta_k/2) \pm (-i/\sqrt{2})b^\dagger_{\delta\mathbf{k},\sigma}(\mathbf{K'})\exp(-i\theta_k/2) \quad (8)$$

which leads to $a^\dagger_{\delta\mathbf{k},\sigma}(\mathbf{K'}) = (-i/\sqrt{2})\exp(-i\theta_k/2)(d^\dagger_{c,\delta\mathbf{k},\sigma}(\mathbf{K'}) + d^\dagger_{v,\delta\mathbf{k},\sigma}(\mathbf{K'}))$, and $b^\dagger_{\delta\mathbf{k},\sigma}(\mathbf{K'}) = (i/\sqrt{2})\exp(i\theta_k/2)(d^\dagger_{c,\delta\mathbf{k},\sigma}(\mathbf{K'}) - d^\dagger_{v,\delta\mathbf{k},\sigma}(\mathbf{K'}))$. It is easy to see that the band operators around **K** and **K'** introduced above anti-commute. Also, around **K** and **K'**, in terms of these band operators, the Hamiltonian (5) is given by $H_0 = \sum_{\delta k} \hbar v_F |\delta k| (d^\dagger_{c,\delta k,\sigma} d_{c,\delta k,\sigma} - d^\dagger_{v,\delta k,\sigma} d_{v,\delta k,\sigma})$. We make the following combination of the band operators in Eqs. (7) and (8):

$$\gamma_{1,A}(\delta k) = d^\dagger_{c,\delta k,\sigma}(K) + i\, d_{v,\delta k,\sigma}(K'); \quad \gamma_{2,A}(\delta k) = d_{c,\delta k,\sigma}(K') - i\, d^\dagger_{v,\delta k,\sigma}(K),$$

$$\gamma_{1,B}(\delta k) = d_{c,\delta k,\sigma}(K) - i\, d^\dagger_{v,\delta k,\sigma}(K'); \quad \gamma_{2,B}(\delta k) = d^\dagger_{c,\delta k,\sigma}(K') + i\, d_{v,\delta k,\sigma}(K).$$

Going back to the fermion operators for A and B sub-lattices, we find that these combinations yield

$$(1/\sqrt{2})(\gamma_{1,A}(\delta k) + i\,\gamma_{2,A}(\delta k)) = (a^\dagger_{\delta k,\sigma}(K) + a_{\delta k,\sigma}(K'))\exp(-i\theta_k/2),$$

$$(1/\sqrt{2})(\gamma_{1,B}(\delta k) + i\,\gamma_{2,B}(\delta k)) = (b^\dagger_{\delta k,\sigma}(K') + b_{\delta k,\sigma}(K))\exp(-i\theta_k/2). \tag{9}$$

It is clear from (9) that if the iso-spin symmetry is completely broken, when $\theta_k = 4\pi, 8\pi, \ldots$ ........for which the Berry phase remains hidden, we have the sub-lattice specific four Dirac particle-hole creation operators $\hat{A}_{ph,\sigma} = (a^\dagger_{0,\sigma} + a_{0,-\sigma})$ and $\hat{C}_{ph,\sigma} = (b^\dagger_{0,\sigma} + b_{0,-\sigma})$, respectively, equal to $(1/\sqrt{2})(\gamma_{1,A,\sigma} + i\,\gamma_{2,A,\sigma})$ and $(1/\sqrt{2})(\gamma_{1,B,\sigma} + i\,\gamma_{2,B,\sigma})$ at the Fermi level, where $\gamma_{1,A,\sigma} = (1/\sqrt{2})(a^\dagger_{0,\sigma} + a_{0,\sigma} + a^\dagger_{0,-\sigma} + a_{0,-\sigma})$, $\gamma_{2,A,\sigma} = (1/i\sqrt{2})(a^\dagger_{0,\sigma} - a_{0,\sigma} - a^\dagger_{0,-\sigma} + a_{0,-\sigma})$, $\gamma_{1,B,\sigma} = (1/\sqrt{2})(b^\dagger_{0,\sigma} + b_{0,\sigma} + b^\dagger_{0,-\sigma} + b_{0,-\sigma})$, $\gamma_{2,B,\sigma} = (1/i\sqrt{2})(b^\dagger_{0,\sigma} - b_{0,\sigma} - b^\dagger_{0,-\sigma} + b_{0,-\sigma})$. The real and imaginary parts of the ordinary fermion operators $\hat{A}_{ph,\sigma}$ and $\hat{C}_{ph,\sigma}$ correspond to eight Majorana fermions as we have $\sigma = \uparrow,\downarrow$ (real spin 'up' and 'down') with $\gamma_{i,\alpha,\sigma} = \gamma^\dagger_{i,\alpha,\sigma}$ (self-conjugate) where $\alpha$ = A/B. The formal manipulations presented above shows that for the Majorana pairs to be realized it is necessary that, apart from the broken iso-spin symmetry, the spin-degeneracy should be lifted. The Dirac operators $\hat{A}_{ph,\sigma}$ and $\hat{C}_{ph,\sigma}$ obey the usual anti-commutation relations: $\{\hat{A}_{ph,\sigma}, \hat{A}^\dagger_{ph,\sigma'}\} = 2\delta_{\sigma\sigma'}$, $\{\hat{C}_{ph,\sigma}, \hat{C}^\dagger_{ph,\sigma'}\} = 2\delta_{\sigma\sigma'}$, $\{\hat{A}_{ph,\sigma}, \hat{A}_{ph,\sigma'}\} = 0$, and $\{\hat{C}_{ph,\sigma}, \hat{C}_{ph,\sigma'}\} = 0$. However, the Majorana operators obey unusual (the product $\gamma_{i,\alpha,\sigma}^2 = 1$ and does not vanish) anti-commutation relations: $\gamma_{i,\alpha,\sigma}\gamma_{j,\beta,\sigma'} + \gamma_{j,\beta,\sigma'}\gamma_{i,\alpha} = 2\delta_{ij}\delta_{\alpha\beta}\delta_{\sigma\sigma'}$. The relevant issue is thus how to lift the (real) spin degeneracy. We shall indicate below that it is possible in the presence of a (tunable) strong Rashba term dominating over the intrinsic spin-orbit coupling.

We write the following general Hamiltonian (H) of the monolayer graphene (MLG) in the basis $(a_{k\uparrow}, b_{k\uparrow}, a_{k\downarrow}, b_{k\downarrow})$ in momentum space involving the Rashba spin-orbit coupling (and the intrinsic spin-orbit coupling)[25,26,27]: mm

$$H = \sum_k (a^\dagger_{k\uparrow}\ b^\dagger_{k\uparrow}\ a^\dagger_{k\downarrow}\ b^\dagger_{k\downarrow})\, \hbar(k) \begin{pmatrix} a_{k\uparrow} \\ b_{k\uparrow} \\ a_{k\downarrow} \\ b_{k\downarrow} \end{pmatrix}, \tag{10}$$

$$\hbar(k) = \begin{pmatrix} -t_{so}\gamma_{so} + M + V & -t\gamma_0 & 0 & t_R(\gamma_{R1} - \gamma_{R2}) \\ -t\gamma^*_0 & t_{so}\gamma_{so} + M - V & t_R(-\gamma^*_{R1} - \gamma^*_{R2}) & 0 \\ 0 & t_R(-\gamma_{R1} - \gamma_{R2}) & t_{so}\gamma_{so} - M + V & -t\gamma_0 \\ t_R(\gamma^*_{R1} - \gamma^*_{R2}) & 0 & -t\gamma^*_0 & -t_{so}\gamma_{so} - M - V \end{pmatrix}$$

$$\tag{11}$$

where the Hamiltonian focuses on the π-orbitals only. The intrinsic spin-orbit coupling(ISOC) term, in coordinate representation, may be written as $H_{so} = (2it_{so}/\sqrt{3}) \sum_{ij} c^\dagger_{i\sigma}(\mathbf{s}\cdot(\mathbf{d}_{kj}\times \mathbf{d}_{ik})) c_{j\sigma}$ where k is connecting the next-nearest neighbor sites i and j; $\mathbf{d}_{kj}$ is a unit lattice vector pointing from site j to site k. Here $c_{i\sigma}$ is π-orbital annihilation operator for an electron with spin σ on site i and **s** are the spin Pauli matrices. The Rashba spin-orbit coupling(RSOC) term, on the other hand, in coordinate representation may be written as $H_R = (it_R) \sum_{ij\mu\nu}[a^\dagger_{i\mu}(s_{\mu\nu}\times \mathbf{d}_{ij})_z b_{j\nu} - h.c]$ where once again **s** are the Pauli matrices representing the electron spin operator and μ, ν = 1, 2 denote the μν matrix elements of the Pauli matrices. The operators $a^\dagger_{i,\sigma}$ and $b^\dagger_{j,\sigma}$, respectively, correspond to the fermion creation operators with real spin σ for A and B sub-lattices in the mono-layer. Upon using the operator $\alpha^\dagger_\mathbf{k} = (1/\sqrt{N}) \sum_i \exp(i\mathbf{k}\cdot\mathbf{R}_i) a^\dagger_{i,\sigma}$ where $\mathbf{R}_i$ is the Bravais vector of the ith unit cell and **k** lies in the first Brillouin zone (and similarly, introducing $\beta^\dagger_\mathbf{k}$ acting on sub-lattice B) it is easy to find that $H_R = (it_R) \sum_{\mathbf{k},\mu\nu}[\alpha^\dagger_{\mathbf{k},\mu} (s_{\mu\nu}\times \mathbf{d}(\mathbf{k}))_z \beta_{\mathbf{k},\nu} - h.c]$ where $\mathbf{d}(\mathbf{k}) = -\sum_{j=1,2,3} \mathbf{d}_j\exp(-i\mathbf{k}\cdot\boldsymbol{\delta}_j)$. For the nearest-neighbor hopping term(t), we have in the same representation $H_0 = -t\sum_{\mathbf{k},\sigma} \gamma_0(\mathbf{k}) \alpha^\dagger_{\mathbf{k},\sigma} \beta_{\mathbf{k},\sigma}$ with $\gamma_0(\mathbf{k}) = \sum_{j=1,2,3} \exp(-i\mathbf{k}\cdot\boldsymbol{\delta}_j)$. The three nearest neighbor vectors are assumed to be $\boldsymbol{\delta}_1 = (a/2)(1,\sqrt{3})$, $\boldsymbol{\delta}_2 = (a/2)(1,-\sqrt{3})$, and $\boldsymbol{\delta}_3 = a(-1,0)$; 'a' is the lattice constant. The second neighbor positions are given by $\mathbf{d}_{1,2} = \pm\mathbf{a}_1$, $\mathbf{d}_{3,4} = \pm\mathbf{a}_2$, $\mathbf{d}_{5,6} = \pm(\mathbf{a}_2 - \mathbf{a}_1)$ where $\mathbf{a}_1 = (a/2)(3,\sqrt{3})$ and $\mathbf{a}_2 = (a/2)(3,-\sqrt{3})$. We consider the term

$$\mathbf{d}(\mathbf{k}) = -\sum_{j=1,2,3} \mathbf{d}_j\exp(-i\mathbf{k}\cdot\boldsymbol{\delta}_j) = -[\mathbf{d}_1\exp(-i\mathbf{k}\cdot\boldsymbol{\delta}_1) + \mathbf{d}_2\exp(-i\mathbf{k}\cdot\boldsymbol{\delta}_2) + \mathbf{d}_3\exp(-i\mathbf{k}\cdot\boldsymbol{\delta}_3)]$$

$$= -(1/2)(\mathbf{i} + \sqrt{3}\,\mathbf{j})[\exp(-ik_x(a/2) - ik_y(\sqrt{3}a/2))] - (1/2)[(\mathbf{i} - \sqrt{3}\,\mathbf{j})[\exp(-ik_x(a/2) + ik_y(\sqrt{3}a/2))]$$

$$+\mathbf{i}\exp(ik_x a).$$

$$= -[(1/2)\{\exp(-ik_x(a/2) - ik_y(\sqrt{3}a/2))\} + (1/2)\{\exp(-ik_x(a/2) + ik_y(\sqrt{3}a/2))\} - \exp(ik_x a)]\,\mathbf{i}$$

$$-[(\sqrt{3}/2)\{\exp(-ik_x(a/2) - ik_y(\sqrt{3}a/2))\} - (\sqrt{3}/2)\{\exp(-ik_x(a/2) + ik_y(\sqrt{3}a/2))\}]\,\mathbf{j}$$

$$= d_1(k)\,\mathbf{i} + d_2(k)\,\mathbf{j}$$

where

$$d_1(k) = -[(1/2)\{\exp(-ik_x(a/2) - ik_y(\sqrt{3}a/2))\} + (1/2)\{\exp(-ik_x(a/2) + ik_y(\sqrt{3}a/2))\} - \exp(ik_x a)],$$

$$d_2(k) = -[(\sqrt{3}/2)\{\exp(-ik_x(a/2) - ik_y(\sqrt{3}a/2))\} - (\sqrt{3}/2)\{\exp(-ik_x(a/2) + ik_y(\sqrt{3}a/2))\}].$$

We now consider the term

$$(s_{\mu\nu} \times \mathbf{d}(\mathbf{k})) = (\sigma_x\,\mathbf{i} + \sigma_y\,\mathbf{j}) \times (d_1(k)\,\mathbf{i} + d_2(k)\,\mathbf{j}) = (\sigma_x d_2(k) - \sigma_y d_1(k))\,\mathbf{k}.$$

With the aid of the right-hand-side we write

$$H_R = (it_R) \sum_{\mathbf{k},\mu\nu}[\alpha^\dagger_{\mathbf{k},\mu}(s_{\mu\nu}\times \mathbf{d}(\mathbf{k}))_z \beta_{\mathbf{k},\nu} - h.c] = (it_R) \sum_\mathbf{k}\left[(\alpha^\dagger_{\mathbf{k}\uparrow}\ \alpha^\dagger_{\mathbf{k}\downarrow})\begin{pmatrix}0 & d_2 + id_1 \\ d_2 - id_1 & 0\end{pmatrix}\begin{pmatrix}\beta_{\mathbf{k}\uparrow}\\ \beta_{\mathbf{k}\downarrow}\end{pmatrix}\right.$$

$$\left.- (\beta^\dagger_{\mathbf{k}\uparrow}\ \beta^\dagger_{\mathbf{k}\downarrow})\begin{pmatrix}0 & d^*_2 + id^*_1 \\ d^*_2 - id^*_1 & 0\end{pmatrix}\begin{pmatrix}\alpha_{\mathbf{k}\uparrow}\\ \alpha_{\mathbf{k}\downarrow}\end{pmatrix}\right]$$

$$= t_R \sum_\mathbf{k}[(i d_2 - d_1)\,\alpha^\dagger_{\mathbf{k}\uparrow}\beta_{\mathbf{k}\downarrow} + (i d_2 + d_1)\,\alpha^\dagger_{\mathbf{k}\downarrow}\beta_{\mathbf{k}\uparrow} + (-i d^*_2 + d^*_1)\,\beta^\dagger_{\mathbf{k}\uparrow}\alpha_{\mathbf{k}\downarrow} + (-i d^*_2 - d^*_1)\,\beta^\dagger_{\mathbf{k}\downarrow}\alpha_{\mathbf{k}\uparrow}].$$

We obtain

$$(i\,d_2 - d_1) = \gamma_{R,1} - \gamma_{R,2},\ (i\,d_2 + d_1) = -\gamma_{R,1} - \gamma_{R,2},$$

$$(-i\,d^*_2 + d^*_1) = -\gamma^*_{R,1} - \gamma^*_{R,2},\ (-i\,d^*_2 - d^*_1) = \gamma^*_{R,1} - \gamma^*_{R,2}.$$

With these paraphernalia we finally obtain

$$\gamma_{R,1} = [\exp(-i\,k_x a/2)\cos(\sqrt{3}k_y a/2) - \exp(i\,k_x a)],\ \gamma_{R,2} = \sqrt{3}\exp(-i\,k_x a/2)\sin(\sqrt{3}k_y a/2),$$

$$\gamma_0 = [2\exp(i k_x a/2)\cos(\sqrt{3}k_y a/2) + \exp(-i k_x a)].$$

In Eq.(11) M, and V, respectively, correspond to the exchange field term, and the staggered AB sub-lattice potential. In the absence of ISOC and the exchange field, the eigenvalues($\varepsilon$) of the Hamiltonian in (11) are given by a bi-quadratic equation in $\varepsilon$:

$$\varepsilon^4 - 2\varepsilon^2\{t^2|\gamma_0|^2 + \tfrac{1}{2}t_R^2(|\gamma_{R,1} + \gamma_{R,2}|^2) + \tfrac{1}{2}t_R^2(|\gamma_{R,1} - \gamma_{R,2}|^2) + V^2\} - f(k_x,k_y) = 0,\quad (12)$$

$$f(k_x,k_y) \equiv [\,t^2 t_R^2\, g(k_x,k_y) - \{V^4 + 2t^2|\gamma_0|^2 V^2 + V^2 t_R^2(|\gamma_{R,1} + \gamma_{R,2}|^2)$$

$$+ V^2 t_R^2(|\gamma_{R,1} - \gamma_{R,2}|^2) + t^4|\gamma_0|^4 + t_R^4(|\gamma_{R,1} + \gamma_{R,2}|^2)(|\gamma_{R,1} - \gamma_{R,2}|^2)\}\,],\quad (13)$$

$$g(k_x,k_y) \equiv [\{16\cos(k_x a/2)\cos^3(\sqrt{3}k_y a/2) + 4\cos(5k_x a/2)\cos(\sqrt{3}k_y a/2) + 8\cos(k_x a)\cos^2(\sqrt{3}k_y a/2)$$

$$+ 24\cos(2k_x a)\sin^2(\sqrt{3}k_y a/2)\cos^2(\sqrt{3}k_y a/2) + 6\cos(k_x a)\sin^2(\sqrt{3}k_y a/2)$$

$$+ 12\cos(k_x a/2)\sin^2(\sqrt{3}k_y a/2)\cos(\sqrt{3}k_y a/2)\}$$

$$-\{8\cos(2k_x a)\cos^4(\sqrt{3}k_y a/2) + 2\cos(k_x a)\cos^2(\sqrt{3}k_y a/2) + 4\cos(k_x a/2)\cos^3(\sqrt{3}k_y a/2)$$

$$+ 8\cos(k_x a)\cos^2(\sqrt{3}k_y a/2) + 2\cos(4k_x a) + 4\cos(5k_x a/2)\cos(\sqrt{3}k_y a/2)\}].\quad (14)$$

It may be noted that if the ISOC and the exchange field terms are included there would be an additional term in (12) involving $\varepsilon$ which makes the eigenvalue equation a quartic. A quartic may be solved for real $\varepsilon$'s by Ferrari method given the suitable choice of the parameters. We note that the RSOC achieves the sought after spin degeneracy lifting: It creates a spin-splitting at the **K**$(2\pi/3a, 2\pi/3\sqrt{3}a)$ and **K'** $(2\pi/3a, -2\pi/3\sqrt{3}a)$ points and the anisotropic band gap

$$G(k_x,k_y) = 2\,[f_0(k_x,k_y) - g_0(k_x,k_y)]^{1/2} \quad (15)$$

between the two bands closer to $\varepsilon = 0$. Here $f_0(k_x,k_y) \equiv [t^2|\gamma_0|^2 + \tfrac{1}{2}t_R^2(|\gamma_{R,1} + \gamma_{R,2}|^2) + \tfrac{1}{2}t_R^2(|\gamma_{R,1} - \gamma_{R,2}|^2) + V^2]$ and $g_0(k_x, k_y) \equiv \sqrt{\{f_0(k_x,k_y)^2 + f(k_x,k_y)\}}$. With the suitable choice of the parameters (such as the relative Rashba coupling strength $(t_R/t) \approx 0.1$, and $(V/t) \approx 0.4$) the term under the radical sign will be positive. In the Figure 1 below, we have contour plotted the gap $G(k_x,k_y)$ on the first Brillouin zone for this strong Rashba coupling. At the **K**$(2\pi/3a, 2\pi/3\sqrt{3}a)$ and **K'** $(2\pi/3a, -2\pi/3\sqrt{3}a)$ points the gap is 0.1384 - nearly the same as the Rashba coupling coupling strength. The discussion above puts a very severe constraint on the appearance of Majoranas, viz. a strong Rashba coupling tunable by a gate voltage.

Our treatment in the next section will not be based on this full four band model of MLG; the complete analysis of the Hamiltonian in (11) and deriving Majorana operators from here is a challenging (future) task. We wish to emphasize that the use of the full four band model, rather than the simplified two band model, is essential to accurately represent the spin-degeneracy lifted nature of the low-energy band structure of MLG and demonstrate convincingly the existence of Majoranas for this system.

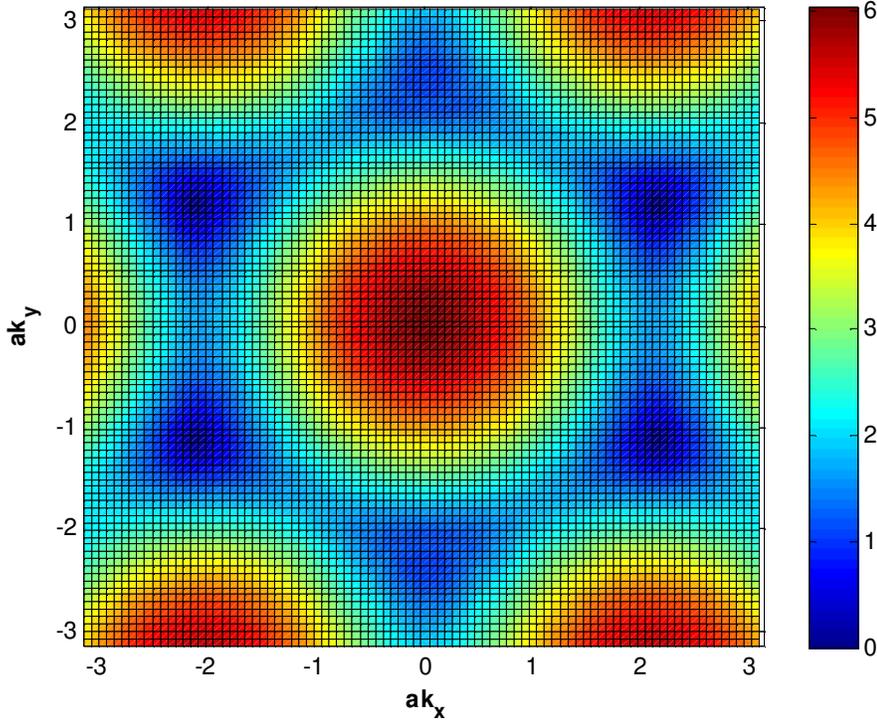

**Figure1.** The contour plot of the gap $G(k_x,k_y)$ on the first Brillouin zone for $(t_R/t)= 0.108$, and $(V/t)=0.41$. At the **K**$(2\pi/3a, 2\pi/3\sqrt{3}a)$ and **K′** $(2\pi/3a, -2\pi/3\sqrt{3}a)$ points the gap is 0.1384.

**4. Longitudinal and Hall conductivities with broken chiral symmetry** In this section we calculate the semi-classical diffusive (longitudinal) conductivity with the broken "chiral" symmetry. We also calculate the Hall conductivity. This symmetry is preserved when a mass term proportional to $\gamma^2\gamma^2$ (see section 2) is added to open a spectral gap. However, this is broken when a mass is introduced via $\gamma^2$ which corresponds to a staggered sub-lattice potential as in the case of Boron nitride. For such massive Dirac fermions, the simplest Dirac Hamiltonian is $H_{K(K'),massive}(a|\delta \mathbf{k}|)$, where

$$H_{K(K'),massive} = \begin{bmatrix} V & \xi \hbar v_F |\delta k| \exp(-i\xi\theta_k) \\ \xi \hbar v_F |\delta k| \exp(i\xi\theta_k) & -V \end{bmatrix}.$$

The index $\xi = \pm$ denotes the two Dirac points **K** and **K′**, respectively. It is easy to see that $[H_{K,massive}, \hat{C}] \neq 0$ and $[H_{K',massive}, \hat{C}^*] \neq 0$. Thus, for the massive fermions corresponding to a staggered sub-lattice potential, the chirality is not a good quantum number. The eigenvalues of the Hamiltonian $H_{K,massive}$ and $H_{K',massive}$ are given by $\pm\sqrt{(V^2 + (\hbar v_F |\delta\mathbf{k}|)^2)}$ leading to the particle-hole symmetry. The eigenstates of the Hamiltonian $H_{K,massive}$ and $H_{K',massive}$, respectively, corresponding to the eigenvalues $\pm\sqrt{(V^2 + (\hbar v_F |\delta\mathbf{k}|)^2)}$ are given by

$$\psi_{\pm, K(K'),massive}(\delta\mathbf{k}) = (1/\sqrt{2}) \begin{pmatrix} \exp(-\frac{i\xi\theta_k}{2}) \\ \pm\xi \exp(\frac{i\xi\theta_k}{2})(\sqrt{(v^2+1)} \pm (-v)) \end{pmatrix} \quad (16)$$

where $v \equiv (V/(\hbar v_F |\delta\mathbf{k}|))$. Applying the Kubo formula [26] to our system we find that the conductivity tensor (including the dynamical Hall conductivity $\sigma_{xy}(\omega)$) is calculated by

$$\sigma_{ii}(\omega) = (ie^2/2\pi\hbar) \int d(\delta\mathbf{k}a) \sum_{\alpha,\beta} [(f(E_\alpha) - f(E_\beta))/(E_\alpha - E_\beta)]$$
$$\times [\{\langle\alpha|\hbar V_i/a|\beta\rangle \langle\beta|\hbar V_i/a|\alpha\rangle\} (E_\alpha - E_\beta - \hbar\omega - i\eta)^{-1}], \quad (17)$$

$$\sigma_{ij,i\neq j}(\omega) = (ie^2/2\pi\hbar) \int d(\delta\mathbf{k}a) \sum_{\alpha,\beta} [(f(E_\alpha) - f(E_\beta))/(E_\alpha - E_\beta)]$$
$$\times [\{\langle\alpha|\hbar V_i/a|\beta\rangle \langle\beta|\hbar V_j/a|\alpha\rangle\} (E_\alpha - E_\beta - \hbar\omega - i\eta)^{-1}] \quad (18)$$

where $f(E_\alpha)$ is the Fermi factor, $\int d(\delta\mathbf{k}a) \to \int (d(\delta k_x a)/2\pi) \int (d(\delta k_y a)/2\pi)$, $E_\alpha$ is the α-th energy eigenvalue and $|\alpha\rangle$ is the corresponding eigenstate. These are the semi-classical diffusive conductivity expressions. Here $V_i$ and $V_j$ are the velocity operators already obtained in section 2. We now make the identifications that the state $|\alpha\rangle$ corresponds to the eigenvalue $(+\sqrt{(V^2+(\hbar v_F|\delta\mathbf{k}|)^2)})$ and the state $|\beta\rangle$ to $(-\sqrt{(V^2+(\hbar v_F|\delta\mathbf{k}|)^2)})$ where, around $\mathbf{K}$,

$$|\alpha\rangle = (1/\sqrt{2}) \begin{pmatrix} \exp(-\frac{i\theta_k}{2}) \\ \exp(\frac{i\theta_k}{2})(\sqrt{(v^2+1)} - v) \end{pmatrix}, |\beta\rangle = (1/\sqrt{2}) \begin{pmatrix} \exp(-\frac{i\theta_k}{2}) \\ -\exp(\frac{i\theta_k}{2})(\sqrt{(v^2+1)} + v) \end{pmatrix}.$$

Around $\mathbf{K'}$, on the other hand, we have $|\alpha\rangle = (1/\sqrt{2}) \begin{pmatrix} \exp(\frac{i\theta_k}{2}) \\ -\exp(-\frac{i\theta_k}{2})(\sqrt{(v^2+1)} - v) \end{pmatrix}$, and

$|\beta\rangle = (1/\sqrt{2}) \begin{pmatrix} \exp(\frac{i\theta_k}{2}) \\ \exp(-\frac{i\theta_k}{2})(\sqrt{(v^2+1)} + v) \end{pmatrix}$. We obtain, around $\mathbf{K}$, $\langle\beta|\hbar V_x/a|\alpha\rangle = (\hbar v_F/a)$ $(-v\cos(\theta_k) + i\sqrt{(v^2+1)}\sin(\theta_k))$, and $\langle\alpha|\hbar V_x/a|\beta\rangle = (\hbar v_F/a)(-v\cos(\theta_k) - i\sqrt{(v^2+1)}\sin(\theta_k))$. This leads to $\{\langle\alpha|\hbar V_x/a|\beta\rangle \langle\beta|\hbar V_x/a|\alpha\rangle\} = [(\hbar v_F/a)^2 \{v^2 + (\delta k_y)^2/|\delta\mathbf{k}|^2\}]$ around $\mathbf{K}$. We obtain the same result around $\mathbf{K'}$. The velocity operator being given by $v_F(\xi\sigma_x, \sigma_y)$ is of no consequence as both the overlap matrix elements above involve $V_x$ only. The swapping, where $|\alpha\rangle$ is identified with the eigenvalue $(-\sqrt{(V^2+(\hbar v_F|\delta\mathbf{k}|)^2)})$ and $|\beta\rangle$ with $(+\sqrt{(V^2+(\hbar v_F|\delta\mathbf{k}|)^2)})$, is also possible in Eq.(17). It will lead to no cancellation in

$[(f(E_\alpha)-f(E_\beta))/(E_\alpha-E_\beta)]$ as a negative sign will be arising out of the denominator in $[(f(E_\alpha)-f(E_\beta))/(E_\alpha-E_\beta)]$ unlike the previous case where the negative sign arises out of the numerator. However, the term $(E_\alpha - E_\beta - \hbar\omega - i\eta)^{-1}$ yields

$$(2\sqrt{(V^2 + (\hbar v_F |\delta\mathbf{k}|)^2)} - \hbar\omega - i\eta)^{-1}$$

in the former case and

$$-(2\sqrt{(V^2 + (\hbar v_F |\delta\mathbf{k}|)^2)} + \hbar\omega + i\eta)^{-1}$$

in the latter case. As we shall see below this will not lead to cancellation. Thus, taking contributions around $\mathbf{K}'$ as well as $\mathbf{K}$ one obtains $\{\langle\alpha|\hbar V_x/a|\beta\rangle\langle\beta|\hbar V_x/a|\alpha\rangle\} = 2[(\hbar v_F/a)^2 \{v^2 + (\delta k_y)^2/|\delta\mathbf{k}|^2\}]$. We shall not consider the mixing of states around $\mathbf{K}$ and $\mathbf{K}'$ in the calculation of the overlap matrix involving velocity operator above. The quantity $[(f(E_\alpha) - f(E_\beta))/(E_\alpha - E_\beta)]$ is equal to $[(-1/2E(\delta\mathbf{k}))\tanh(\beta E(\delta\mathbf{k})/2)]$ where $\beta = (k_B T)^{-1}$ and $E(\delta\mathbf{k}) = \sqrt{(V^2 + (\hbar v_F|\delta\mathbf{k}|)^2)}$. Finally, using the result $(x \pm i\eta)^{-1} = [P(x^{-1}) \pm (1/i)\pi\delta(x)]$ where P represents a Cauchy's principal value, we are led to

$$(E_\alpha - E_\beta - \hbar\omega - i\eta)^{-1} = P[(E_\alpha - E_\beta - \hbar\omega)^{-1}] + i\pi\delta(E_\alpha - E_\beta - \hbar\omega)$$

$$= P[(2E(\delta\mathbf{k}) - \hbar\omega)^{-1}] + i\pi\delta(2E(\delta\mathbf{k}) - \hbar\omega).$$

Also,

$$-(2E(\delta\mathbf{k}) + \hbar\omega + i\eta)^{-1} = -P[(2E(\delta\mathbf{k}) + \hbar\omega)^{-1}] + i\pi\delta(2E(\delta\mathbf{k}) + \hbar\omega).$$

Upon substituting these results in Eq.(17), and multiplying by a factor 2 due to the spin degeneracy, we find that the imaginary part 'Im $\sigma_{xx}(\omega)$' goes to zero. The reason being this could be written as

$$\text{Im } \sigma_{xx}(\omega) = (4e^2/i h) \int_{-\infty}^{+\infty} d\varepsilon \int d(\delta\mathbf{k}a)[(2\varepsilon)^{-1}\tanh(\beta\varepsilon/2)] [(\hbar v_F/a)^2 \{v^2 + (\delta k_y)^2/|\delta\mathbf{k}|^2\}]$$

$$[\delta(\varepsilon - E(\delta\mathbf{k})) + \delta(\varepsilon + E(\delta\mathbf{k}))][P\{(2\varepsilon - \hbar\omega)^{-1} - (2\varepsilon + \hbar\omega)^{-1}\}]$$

upon introducing a $\delta$-function density of state $[\delta(\varepsilon - E(\delta\mathbf{k})) + \delta(\varepsilon + E(\delta\mathbf{k}))]$. This possesses peaks at $\pm E(\delta\mathbf{k})$ and the integrand

$$I_1(\varepsilon) = [(2\varepsilon)^{-1}\tanh(\beta\varepsilon/2)][\delta(\varepsilon - E(\delta\mathbf{k})) + \delta(\varepsilon + E(\delta\mathbf{k}))][P\{(2\varepsilon - \hbar\omega)^{-1} - (2\varepsilon + \hbar\omega)^{-1}\}]$$

of the integral $\int_{-\infty}^{+\infty} d\varepsilon\, I_1(\varepsilon)$ above is an odd function of $\varepsilon$. We, thus, obtain

$$\sigma_{xx}(\omega) = (4\pi e^2/h) \int_{-\infty}^{+\infty} d\varepsilon \int d(\delta\mathbf{k}a)[(2\varepsilon)^{-1}\tanh(\beta\varepsilon/2)][(\hbar v_F/a)^2\{v^2 + (\delta k_y)^2/|\delta\mathbf{k}|^2\}]$$

$$[\delta(\varepsilon - E(\delta\mathbf{k})) + \delta(\varepsilon + E(\delta\mathbf{k}))]$$

$$\times [\delta(2E(\delta\mathbf{k}) - \hbar\omega) + \delta(2E(\delta\mathbf{k}) + \hbar\omega)]. \qquad (19)$$

The integral $\int_{-\infty}^{+\infty} d\varepsilon\, [(2\varepsilon)^{-1}\tanh(\beta\varepsilon/2)][\delta(\varepsilon - E(\delta\mathbf{k})) + \delta(\varepsilon + E(\delta\mathbf{k}))]$ is trivial and equals $[E(\delta\mathbf{k})^{-1}\tanh(\beta E(\delta\mathbf{k})/2)]$. Therefore

$$\sigma_{xx}(\omega) = (4\pi\, e^2/h) \int d(\delta\mathbf{k}a)\, [E(\delta\mathbf{k})^{-1} \tanh(\beta\, E(\delta\mathbf{k})/2)]\, [(\hbar v_F/a)^2 \{v^2 + (\delta k_y)^2/|\delta\mathbf{k}|^2\}]$$

$$\times\, [\delta(2\, E(\delta\mathbf{k}) - \hbar\omega) + \delta(2\, E(\delta\mathbf{k}) + \hbar\omega)]. \quad (20)$$

We notice that both ac and dc longitudinal conductivities could be directly calculated by this formula. Also, the ac longitudinal conductivity for the pristine pure graphene has δ-function peaks at $\omega = \pm 2E(\delta\mathbf{k})/\hbar$. At this point we note that a real graphene system is always disordered. Introducing the effect of disorder in an ad-hoc manner through a level-broadening factor $\eta = \gamma\,(\hbar v_F/a)$, where $\gamma = (\tau v_F\, k_F)^{-1}$, $k_F$ is the Fermi momentum related to the carrier density $n_e$ by $k_F = (\pi\, n_e)^{1/2}$ ($n_e$ can be controlled by the application of a back-gate voltage after transferring the graphene sheet to a dielectric substrate), and τ is the impurity scattering time (the mean free path $\ell = \tau\, v_F$), we may write

$$\sigma_{xx}(\omega) = (4\pi\, e^2/h) \int d(\delta\mathbf{k}a) [\tanh(\beta\,(\hbar v'_F a^{-1})(a|\delta\mathbf{k}|)/2)] (1/|a\delta\mathbf{k}|\sqrt{v^2 + 1})$$

$$\times\, [(\hbar v_F\, a^{-1})\{v^2 + (a\delta k_y)^2/|a\delta\mathbf{k}|^2\}]$$

$$\times [\{\eta/(\eta^2 + (2\,\hbar v'_F a^{-1}(a|\delta\mathbf{k}|) - \hbar\omega)^2)\} + \{\eta/(\eta^2 + (2\,\hbar v'_F a^{-1}(a|\delta\mathbf{k}|) + \hbar\omega)^2)\}]$$
$$(21)$$

where the modified Fermi velocity $\hbar v'_F \approx (\sqrt{3}a|t|/2)\sqrt{(v^2 + 1)}$.

For the dc case ($\omega \to 0$), we obtain the semi-classical diffusive (longitudinal) conductivity as

$$\sigma_{xx}(0) = (8\pi e^2/h) \int d(\delta\mathbf{k}a)[\tanh(\beta\,(\hbar v'_F a^{-1})(a|\delta\mathbf{k}|)/2)]\, (1/|a\delta\mathbf{k}|\sqrt{v^2 + 1})$$

$$\times\, [(\hbar v_F\, a^{-1})\{v^2 + (a\delta k_y)^2/|a\delta\mathbf{k}|^2\}] \times [\eta/(\eta^2 + (2\,\hbar v'_F a^{-1}(a|\delta\mathbf{k}|))^2)]. \quad (22)$$

The integral $\int d(\delta\mathbf{k}a) \to {}_{-\iota\pi}\!\int^{+\iota\pi}(d(\delta k_x a)/2\pi\, {}_{-\pi\iota}\!\int^{+\pi\iota}(d(\delta k_y a)/2\pi$ where $\iota \ll 1$ to ensure that we are not far away from the Dirac points. Since the integrand in Eq.(22) is even function of $(\delta k_x a)$ and $(\delta k_y a)$, the integral would be non-zero. For the (δ**k**)-integration purpose in (22), we first divide the a small region of the momentum space centered at Dirac point into finite number of rectangular patches. We next determine the numerical values corresponding to each of these patches of the momentum-dependent dc conductivity density and sum these values. We have generated these values through the surface plot using the 'MATLAB' package. The conductivity could be calculated at a finite temperature. At the room temperature T = 300 K, the momentum summation is found to be nearly 0.0803 for $(aV/\hbar v_F) = 0.1$ and $\gamma = 10^{-4}$. We have assumed $n_e \sim 10^{18}\,m^{-2}$, the impurity concentration $n_{imp} \sim 10^{15}\,m^{-2}$, and $\ell \sim 10\mu m$ which yield $k_F\ell \gg 1$ as is required in a Boltzmann theory of transport. All these lead to the semi-classical diffusive (longitudinal) conductivity being nearly $(2.0182 e^2/h)$ at room temperature for the disordered system. Evidently, this is an overestimation (see section 5). Never-the-less, we have been able to qualitatively capture the fact that the room temperature conductivity of graphene is finite and the contribution to the conductivity arises from the momentum very close to the Dirac points. Furthermore, even when the chiral symmetry remains preserved, that is V = 0, the conductivity does not vanish. For cleaner samples($n_{imp} < 10^{15} m^{-2}$), the mean free path($\ell$) will be higher and as a result $\eta < 10^{-4}$ and one gets closer to the universal value ($4e^2/\pi h$)[28].

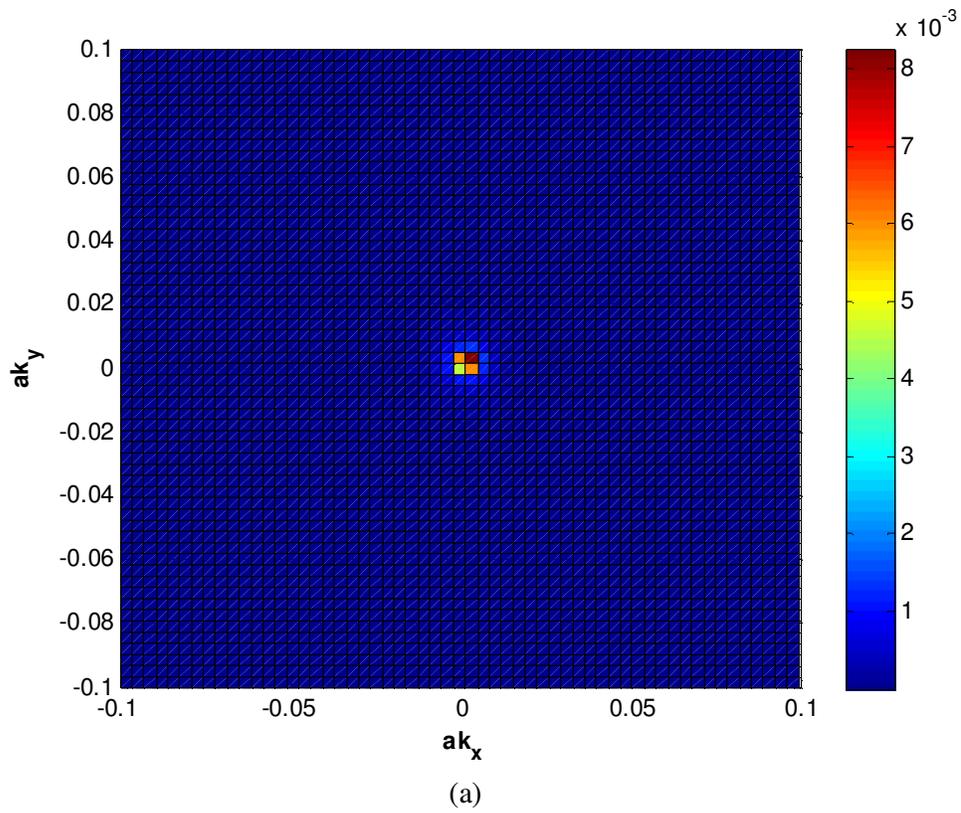

(a)

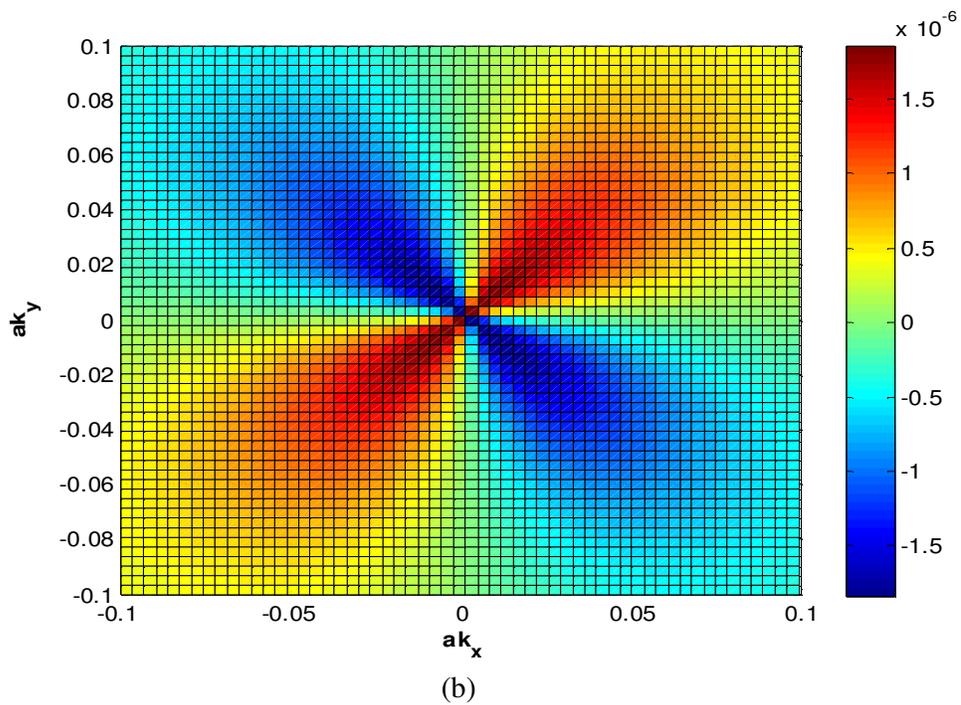

(b)

**Figure 2.** (a) The contour plot of the longitudinal conductivity density as a function of $(k_x, k_y)$ close to the Dirac point for $(aV/\hbar v_F) = 0.1$, $\gamma = 10^{-4}$, and T = 300K. The contribution to the conductivity arises from the momentum very close to the Dirac points only. (b) The contour plot of the Hall conductivity density as a function of $(k_x, k_y)$ close to the Dirac point for $(aV/\hbar v_F) = 0.1$, $\gamma = 10^{-4}$, and T = 300K.

The dynamical Hall conductivity $\sigma_{xy}(\omega)$ is calculated by Eq.(18). Around **K** and **K′**, respectively, the matrix element $\{i\langle\alpha|\hbar V_x/a|\beta\rangle \langle\beta|\hbar V_y/a|\alpha\rangle\}$ is equal to $[(\hbar v_F/a)^2 \{-i(\delta k_x \delta k_y)/|\delta \mathbf{k}|^2 - V\sqrt{V^2+1}\}]$ and $[(\hbar v_F/a)^2\{-i(\delta k_x \delta k_y)/|\delta \mathbf{k}|^2 + V\sqrt{V^2+1}\}]$ with the eigenstates above. The aforementioned 'swapping of the identifications' is also possible. Thus, taking contributions around **K′** as well as **K**, one obtains $\{i\langle\alpha|\hbar V_x/a|\beta\rangle \langle\beta|\hbar V_y/a|\alpha\rangle\}=2[(\hbar v_F/a)^2 \{-i(\delta k_x \delta k_y)/|\delta \mathbf{k}|^2\}]$. We do not consider the mixing of states around **K** and **K′** in the calculation of the overlap matrix involving velocity operator for the Hall conductivity as well. As in Eqs.(21) and (22), for non-vanishing V, we may write

$$\sigma_{xy}(0) = (8\pi e^2/h) \int d(\delta \mathbf{k} a)[\tanh(\beta (\hbar v'_F a^{-1})(a|\delta \mathbf{k}|)/2)] (1/|a\delta \mathbf{k}|\sqrt{v^2+1})$$

$$\times [(\hbar v_F a^{-1}) \{(a\delta k_x)(a\delta k_y)/|a\delta \mathbf{k}|^2\}] \times [\eta/(\eta^2 + (2\hbar v'_F a^{-1}(a|\delta \mathbf{k}|))^2)] \qquad (23)$$

where the modified Fermi velocity $\hbar v'_F \approx (\sqrt{3}a|t|/2)\sqrt{(v^2+1)}$. The integral $\int d(\delta \mathbf{k} a) \to {}_{-\iota\pi}\int^{+\iota\pi}(d(\delta k_x a)/2\pi {}_{-\pi\iota}\int^{+\pi\iota}(d(\delta k_y a)/2\pi$, where $\iota \ll 1$ to ensure that we are not far away from the Dirac points as before. On account of the presence of $(\delta k_x a) \times (\delta k_y a)$ in the integrand, the integral turns out to be vanishingly small (but not zero) as could be seen from the color bar of the contour plot of the Hall conductivity density shown in Figure 2(b). Thus, one sees that the " broken chiral symmetry" is at the heart of the non-zero Hall conductivity when magnetic field (B) is zero. In order to have higher value of the Hall conductivity in the absence of the magnetic field, it seems necessary to consider the Hamiltonian (see Eq.(11)) in section 3 involving the broken chiral symmetry and the Rashba spin-orbit coupling.

**5. Concluding remarks** In section 3 we have seen that around **K** and **K′** the diagonalized Hamiltonian $H_0$, in terms of the band operators together with a staggered a potential $(m v_F^2)$ which takes on different values on the two sub-lattices, appears as $\sum_{\delta \mathbf{k}} \hbar v_F |\delta \mathbf{k}| (d^\dagger_{c,\delta k,\sigma} d_{c,\delta k,\sigma} - d^\dagger_{v,\delta k,\sigma} d_{v,\delta k,\sigma}) + \sum_{\delta \mathbf{k}} m v_F^2 (d^\dagger_{c,\delta k,\sigma} d_{c,\delta k,\sigma} + d^\dagger_{v,\delta k,\sigma} d_{v,\delta k,\sigma})$. Since a momentum ($\delta \mathbf{k}$) state in 'c' and 'v' can either be associated with the valley **K** (with probability amplitudes, respectively, as $\alpha_K$ and $\beta_K$) or with the valley **K′** (with probability amplitudes, respectively, as $\alpha_{K'}$ and $\beta_{K'}$), one may write the most general momentum ($\delta \mathbf{k}$) state in 'c' as $|\Psi(\delta \mathbf{k})\rangle_c = (\alpha_K \psi_{c,K}(\delta \mathbf{k}) + \alpha_{K'} \psi_{c,K'}(\delta \mathbf{k}))$. Similarly, the most general momentum ($\delta \mathbf{k}$) state in 'v' is $|\Psi(\delta \mathbf{k})\rangle_v = (\beta_K \psi_{v,K}(\delta \mathbf{k}) + \beta_{K'} \psi_{v,K'}(\delta \mathbf{k}))$. A real-space spinorial state $\psi(\mathbf{r})$ is given by $\psi(\mathbf{r}) = (\Omega_{BZ})^{-1} \int \delta^2(\delta \mathbf{k}) \exp(i\delta \mathbf{k} \cdot \mathbf{r})[|\Psi(\delta \mathbf{k})\rangle_c + |\Psi(\delta \mathbf{k})\rangle_v]$. Thus, in order to localize graphene electrons on a single sub-lattice (A or B), one needs to superpose states at different energy, that is states from the valence band and from the conduction band, which arise from different valleys. Therefore each electronic wave function at fixed energy has components on both sub-lattices, possibly with equal weight. The only exception are states exactly at zero energy in the presence of a perpendicular magnetic field - their number increases to macroscopic size when a magnetic field is applied. In this case (the zero-energy level), states in the **K** valley reside on one sub-lattice (say B) and those in **K'** on the other one. The problem of graphene electrons localization mechanism on a single sub-lattice, other than that corresponding to the zero-energy level in the presence of a magnetic field, needs a serious investigation which we wish to take up in a future communication.

We emphasize that the non-zero value of the dc conductivity obtained here is based on the Kubo formula which essentially leans on the Boltzmann equation. The equation becomes tractable in the relaxation time approximation. We point out that the important aspect of our description is to introduce the semi-classical diffusive mechanism which is applicable in all disordered graphene samples. It leads to an approximate, possibly non-universal, minimum

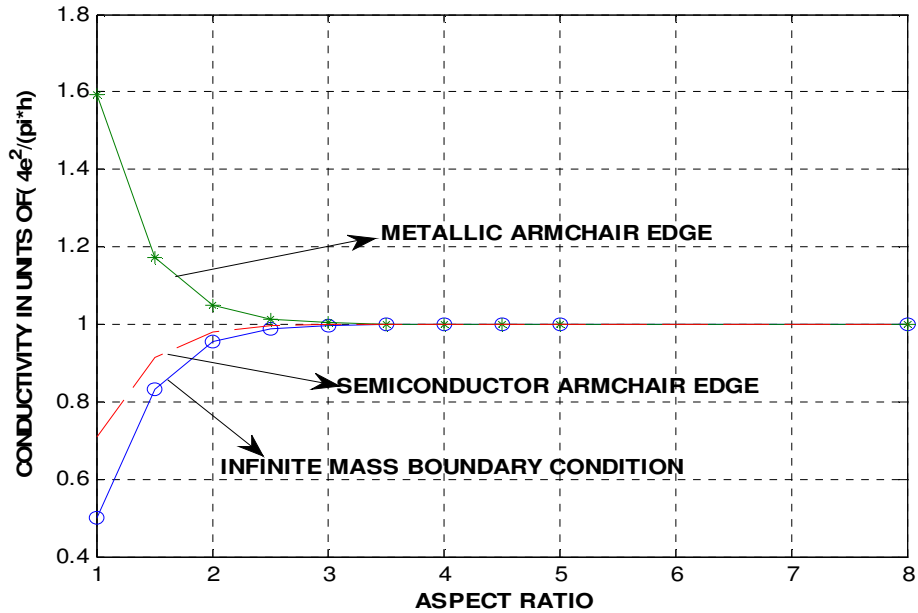

(a)

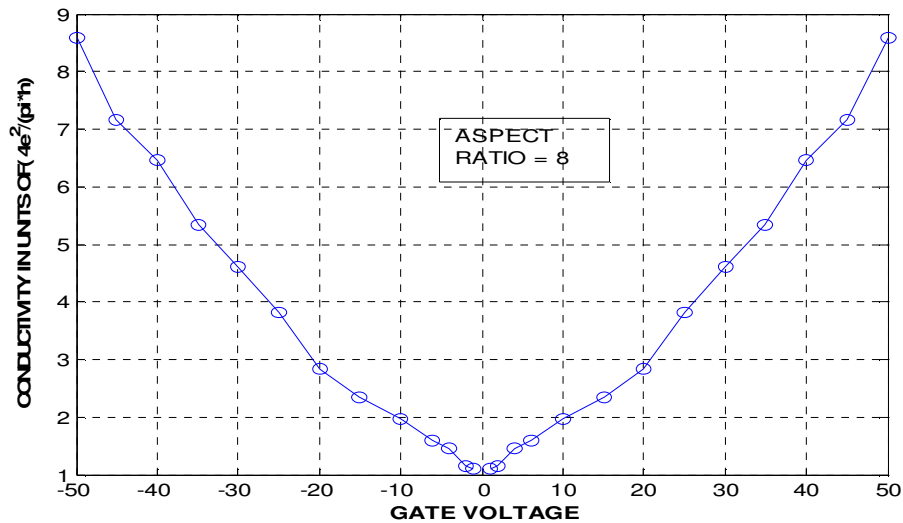

(b)

**Figure 3.** (a)The conductivity ($\sigma$) at the Dirac point ($V_{gate} = 0$), as a function of the aspect ratio(AR) of the graphene strip. For AR ~ 4, through the curves corresponding to the 'infinite mass boundary condition', 'semiconductor armchair edge', and 'metallic armchair edge', we see that the conductivity $\sigma$ converges at ($4e^2/\pi$

h). (b) The conductivity (σ), for the aspect ratio = 8.0, as a function of the non-zero gate voltage($V_{gate}$), i.e. away from the Dirac point. The curve corresponds to the 'infinite mass boundary condition'. The minimal conductivity at the Dirac point in this case is ($4e^2/\pi h$).

graphene conductivity at low induced carrier density. In Figure 3 we have shown the plots of the conductivity as a function of the aspect ratio of graphene strip and the gate voltage, for the sake of completeness, obtainable through the results of the authors in ref.[28] where they discuss the graphene's ''ballistic conductivity.'' One obviously has to accept the "tunneling mechanisms" [28] that leads to a "universal" minimum intrinsic graphene conductivity at the Dirac point in the clean limit and near zero temperature. Within the scope of our present description, there seems to be no way of reconciliation with this exceptional result.

We note that our scheme to realize Majorana fermions by using spin-orbit interaction (SO) (and Zeeman magnetic field) is not entirely a novel one. In a different situation (a BCS s-wave super-fluid of ultra-cold fermionic atoms in an optical lattice with a laser-field-generated effective SO interaction) was considered for the first time by Sato et al.[6]. These authors have derived an important condition $h > \sqrt{(\mu^2 + \delta^2)}$ for the Majorana fermion for the first time, where h is the Zeeman field, μ is the chemical potential, and δ is an s-wave gap function. The same scheme was subsequently considered by Sau et al.[7] in a different setting. Actually the model Hamiltonian of Sau et al.[7] is identical to that in the second work in reference [6]. Under the condition $h > \sqrt{(\mu^2 + \delta^2)}$, it was shown by Sato et al.[6] that the bulk topological number of the system considered becomes nonzero, a topologically protected Majorana edge mode appears, and a Majorana zero mode exists in a vortex core. Our system being different, we do not tread this path. The first step for us in future is to obtain a low-energy Dirac Hamiltonian from Eq. (11) and calculate the energy eigenvalues and the corresponding eigenvectors. With the aid of the latter, one hopes to arrive at the Majorana-like operators for graphene in the second quantization language including the effect of atomically sharp scatterers in a rigorous manner.